\documentclass[journal,twoside]{IEEEtran}
\usepackage{graphicx,amsmath,amssymb}
\usepackage{verbatim}
\usepackage[noadjust]{cite}
\usepackage{array}
\usepackage{cases}
\usepackage{url}
\usepackage{multirow}
\newtheorem{theorem}{Theorem}[section]
\newtheorem{proposition}[theorem]{Proposition}

\begin{document}
\title{Self-Synchronizing Pulse Position Modulation with Error Tolerance}
\author{Yuichiro~Fujiwara,~\IEEEmembership{Member,~IEEE}%
\thanks{This work was supported by JSPS.}%
\thanks{The author is with the Division of Physics, Mathematics and Astronomy, California Institute of Technology, MC 253-37, Pasadena, CA 91125 USA
{(email: yuichiro.fujiwara@caltech.edu)}.}%
\thanks{Copyright \copyright\ 2013 IEEE. Personal use of this material is permitted.
However, permission to use this material for any other purposes must be obtained from the IEEE by sending a request to pubs-permissions@ieee.org.}}
\markboth{IEEE transactions on Information Theory,~Vol.~x, No.~xx,~month~year}
{Fujiwara: Self-synchronizing pulse position modulation with error tolerance}


\maketitle

\begin{abstract}
Pulse position modulation (PPM) is a popular signal modulation technique which converts signals into $M$-ary data
by means of the position of a pulse within a time interval.
While PPM and its variations have great advantages in many contexts,
this type of modulation is vulnerable to loss of synchronization,
potentially causing a severe error floor or throughput penalty even when little or no noise is assumed.
Another disadvantage is that this type of modulation typically offers no error correction mechanism on its own,
making them sensitive to intersymbol interference and environmental noise.
In this paper we propose a coding theoretic variation of PPM
that allows for significantly more efficient symbol and frame synchronization as well as strong error correction.
The proposed scheme can be divided into a synchronization layer and a modulation layer.
This makes our technique compatible with major existing techniques such as standard PPM, multipulse PPM, and expurgated PPM as well
in that the scheme can be realized by adding a simple synchronization layer to one of these standard techniques.
We also develop a generalization of expurgated PPM suited for the modulation layer of the proposed self-synchronizing modulation scheme.
This generalized PPM can also be used as stand-alone error-correcting PPM with a larger number of available symbols.
\end{abstract}

\begin{IEEEkeywords}
Pulse position modulation, PPM, synchronization, error correction, self-synchronizing code, comma-free code, combinatorial design,
optical orthogonal code.
\end{IEEEkeywords}

\IEEEpeerreviewmaketitle

\section{Introduction}
\IEEEPARstart{T}{his} work addresses the problem of symbol and frame synchronization during signal modulation
by means of the position of each pulse in the time domain when the channel is assumed to be noisy.
The approach taken here is to rethink the problem in a coding theoretic framework
and give a theoretical and general foundation.

Among various forms of information, perhaps the simplest of them is binary information.
In many communications scenarios, however, it is beneficial to transmit data in $M$-ary format with $M > 2$.
For this reason, there have been proposed various modulation techniques that support not only binary format but also $M$-ary format with large $M$.

\textit{Pulse position modulation}, or \textit{PPM} for short, is one of the more popular modulation techniques \cite{P}.
In this modulation technique, each symbol occupies a time interval of equal length.
A symbol interval is divided into $Q$ time slots of equal length, where exactly one pulse is transmitted at one of the $Q$ time slots.
Which symbol each interval represents is determined by at which time slot the unique pulse is transmitted.
Because there are $Q$ choices of pulse positions for each symbol interval, PPM offers $M = Q$ distinct symbols.

While PPM is one of the most fundamental forms of signal modulation in use today, there are some inherent drawbacks.
For instance, in a communications system with a severe peak power constraint,
PPM becomes inefficient as the number of symbols increases
because the energy per symbol drops accordingly.

\textit{Multipulse} PPM is a generalization of PPM to mitigate this problem,
where $K$ pulses are sent during each symbol interval by using $K$ out of $Q$ time slots so that $M= {{Q}\choose{K}}$ symbols can be represented \cite{SN}.
Let $\mathcal{P}_K = \left\{\boldsymbol{v}_i \in \mathbb{F}_2^Q \ \middle\vert\ \operatorname{wt}(\boldsymbol{v}_i) = K\right\}$
be the set of all $Q$-dimensional binary vectors $\boldsymbol{v}_i$ of weight $K$.
We let $1$s represent time slots at which single pulses are transmitted and $0$s those at which no pulse is sent.
From the viewpoint of coding theory,
the symbols of PPM can be seen as the binary constant-weight code $\mathcal{P}_1$ of length $Q$, weight one,
and minimum distance two with $Q$ codewords,
whereas the symbols of multipulse PPM can be regarded as the binary constant-weight code $\mathcal{P}_K$
of length $Q$, weight $K$, and minimum distance two with $ {{Q}\choose{K}}$ codewords.

The noncoherent nature of PPM and multipulse PPM makes them attractive to communications systems
in which coherent detection is expensive or impossible, such as optical communications systems \cite{opticalcommunications}.
However, as is evident from the fact that the codes $\mathcal{P}_1$ and $\mathcal{P}_K$ are both of minimum distance two,
one of the major disadvantages of these modulation techniques is that
they are inherently vulnerable to intersymbol interference and natural environmental noise.
Hence, PPM and multipulse PPM require strong error correcting schemes at a higher level when errors are of concern.

\textit{Expurgated} PPM is a recently proposed modulation technique that generalizes PPM
in a way error correction can be provided at the modulation stage while offering the same number $M = Q$ of symbols \cite{NB}.
If we see it from the coding theoretic point of view,
the key idea of expurgated PPM can be understood as using special combinatorial designs called
\textit{symmetric} \textit{designs} to define constant-weight codes with large minimum distances
which allow for simple implementation.

The above three modulation techniques still share the other major disadvantage, namely the susceptibility to loss of synchronization.
As in earlier research on synchronization for this type of modulation \cite{G,G2,PG,CG,VG2,VG},
we assume that slot synchronization is always provided so that the magnitude of misalignment can by expressed by a multiple of the length of a time slot.
Under this assumption, erroneous symbol synchronization means that
the window of the receiver is aligned to the consecutive $Q$ time slots that
consist of the last $i$ time slots of one symbol interval and the first $Q-i$ time slots of the following one for some positive integer $i \leq Q-1$.
Erroneous frame synchronization is understood the same way by regarding a set of consecutive $f$ symbol intervals as a frame of length $f$ with $fQ$ time slots.

In PPM, even if the channel is completely noiseless, erroneous symbol synchronization can not be detected
if there happens to be exactly one pulse within the misaligned window.
Similarly, the receiver can not detect erroneous symbol synchronization under multipulse PPM
if there are exactly $K$ pulses within the misaligned window.
In the case of expurgated PPM, erroneous symbol synchronization will go unnoticed if the $Q$ slots in the misaligned window
form a valid codeword or invalid one within the decodable range.
Hence, without some sort of synchronization mechanism, these modulation techniques may exhibit a severe error floor.

The known method for alleviating this synchronization problem in the literature is
to periodically insert a synchronization marker that consists of $sQ$ time slots $T_i$, $0 \leq i \leq sQ-1$, for some positive integer $s$
in which for any nonnegative integer $j \leq s-1$ the $Q$ consecutive time slots
$T_{i+jQ}$, $0 \leq i \leq Q-1$, form a valid codeword (see, for example, \cite{G,G2,VG}).
In other words, a certain pattern of consecutive $s$ symbols is periodically inserted to signal the boundaries.
Ideally, the off-peak autocorrelations of the synchronization marker should be as small as possible to suppress
the probability that the receiver misses the marker or is deceived by a false one at an unintended position.
For PPM and multipulse PPM, it is possible to find a synchronization marker that causes no ambiguity as long as the channel is noiseless.
However, an ambiguous synchronization marker can not be short.
If the channel is noisy, synchronization with this approach becomes increasingly difficult and complicated as well.

The primary purpose of this paper is to provide a unified solution to both error tolerance and synchronization by using coding theory.
We propose a modulation scheme in which reliable symbol and frame synchronization can be achieved under the presence of noise
through the same type of correlation receiver used for expurgated PPM.
The required overhead in terms of the size of a synchronization marker is significantly smaller than the known method.
The signal modulation can be performed separately, so that
the known modulation techniques of PPM kind can be exploited straightforwardly.
We also develop a generalized format of expurgated PPM which has a larger number of symbols and increased minimum distance than
standard PPM and expurgated PPM.
This generalized PPM can also be used for the proposed self-synchronizing modulation scheme
in place of other modulation techniques of PPM kind to offer higher error tolerance and/or better throughput.

\section{The scheme}
We describe our scheme as a special class of constant-weight codes by exploiting the coding theoretic view introduced in the previous section.
For the sake of generality, a simple setting is assumed
where the probability that the receiver fails to correctly decode symbols decreases monotonously as the minimum distance increases.
Thus, for the most part, our framework will focus on the minimum distance
and generally aim for the largest possible codewords for given parameters.
A more detailed analysis of the performance of our proposed modulation technique is briefly discussed at the end of this paper.

This section is divided into two subsections. Subsection \ref{sync} is devoted to our self-synchronization mechanism
that allows for separate implementation of a modulation layer based on pulse positions in the time domain.
Subsection \ref{mod} gives a generalized version of expurgated PPM that can be used both as a stand-alone modulation scheme
with error correction and as the modulation part of our scheme.

\subsection{Synchronization layer}\label{sync}
As is pointed out in the previous section,
PPM and its variations can be interpreted in terms of constant-weight codes.
In this coding theoretic framework,
the number of codewords corresponds to the number of available symbols while
the minimum distance corresponds to the error tolerance capability.
The constraint that the weight of a codeword is constant ensures equal energy across symbols.
Hence, a good signal modulation technique based on the pulse positions within symbol intervals of fixed length
corresponds to a constant-weight code of large minimum distance with many codewords that can be decoded in a certain simple manner.

Since this coding theoretic view offers a clear picture of how the sensitivity to noise may be alleviated,
it would be natural to ask if it is also possible to exploit the framework to overcome the other major weakness, namely the vulnerability to loss of synchronization,
while preserving other major features of existing standard techniques.
This subsection answers this question in the affirmative
by developing a theory of symbol and frame synchronization for signal modulation of PPM type.
In what follows, we use binary codes to represent symbols expressed by pulse positions,
where $0$s in a codeword correspond to time slots with no pulse
while $1$s represent those at which single pulses are sent.

A \textit{self-synchronizing code} $\mathcal{C} \subset \mathbb{F}_2^n$ is a binary block code of length $n$ where the 
symbol string formed by an overlapped portion of any two concatenated codewords is not a valid codeword.
In the coding theory literature, self-synchronizing codes are also called \textit{comma-free codes}.
The property that no codeword appears as a substring of two adjacent codewords allows for block synchronization
without any external help as long as synchronization is provided at the bit level.

The key idea of our approach is to use a special self-synchronizing code
that allows for synchronization by observing only part of the window and frees up the rest for modulation.
To obtain constant-weight self-synchronizing codes with desirable properties for our purpose,
we employ combinatorial design theory.

Take a sequence of codewords of a binary block code of length $n$.
A {\it splice} of length $n$ between codeword $\boldsymbol{x} = (x_0, x_1, \dots, x_{n-1})$
and the following codeword $\boldsymbol{y} = (y_0, y_1, \dots, y_{n-1})$ in the codeword sequence is
a concatenated binary sequence $(x_{n-i}, \dots, x_{n-1}, y_0, \dots y_{n-i-1})$ composed of the last $i$ bits of $\boldsymbol{x}$
and the first $n-i$ bits of $\boldsymbol{y}$ for some positive integer $i \leq n-1$.

A binary block code of length $n$ is said to be of \textit{comma-free index} $\rho$ if the Hamming distance between any codeword $\boldsymbol{z}$
and any splice of length $n$ between any two codewords $\boldsymbol{x}, \boldsymbol{y}$ is at least $\rho$.
By definition, a self-synchronizing code is a binary block code of comma-free index at least $1$.
It is straightforward to see that with a hard-decision algorithm a self-synchronizing code of comma-free index $\rho$
assures block synchronization under the presence of up to $\lfloor\frac{\rho-1}{2}\rfloor$ bit flips (or errors) in the received message of length $n$.
In an additive white Gaussian noise (AWGN) channel, for example, one may use a correlation receiver for soft-decision synchronization
to take advantage of the Hamming distance between a valid codeword and a splice.

A \textit{difference system of sets} (DSS) of \textit{index} $\rho$ over ${\textit{\textbf{Z}}}_n$
is a family of disjoint subsets $D_i$ of ${\textit{\textbf{Z}}}_n$ such that
the multi-set 
\begin{equation}
\label{equ}
\{ a-b \pmod{n} \ \vert \ a \in D_i, b \in D_j, i \not=j \}
\end{equation}
 contains
every $d \in {\textit{\textbf{Z}}}_n\setminus \{0\}$ at least $\rho$ times.
The difference between two elements from different subsets of ${\textit{\textbf{Z}}}_n$ is called an \textit{outer difference}.
A DSS is {\it perfect} if the multi-set defined in (\ref{equ}) contains
every $d \in {\textit{\textbf{Z}}}_n\setminus \{0\}$ exactly $\rho$ times.
A DSS is \textit{regular} if all subsets $D_i$ are of the same size.
For instance, the set $\{\{1,2\}, \{3,5\}\}$ over ${\textit{\textbf{Z}}}_8$ forms a regular DSS of index one
because every nonzero outer difference appears at least once as follows:
\begin{align*}
1 - 3 &\equiv 6 \pmod{8}, \ 3 - 1 \equiv 2 \pmod{8},\\
1 - 5 &\equiv 4 \pmod{8}, \ 5 - 1 \equiv 4 \pmod{8},\\
2 - 3 &\equiv 7 \pmod{8}, \ 3 - 2 \equiv 1 \pmod{8},\\
2 - 5 &\equiv 5 \pmod{8}, \ 5 - 2 \equiv 3 \pmod{8}.
\end{align*}
This DSS is not perfect because $4$ appears twice as an outer difference
while each of the other elements of ${\textit{\textbf{Z}}}_8\setminus \{0\}$ occurs exactly once.

The original motivation of the study of DSSs was to realize self-synchronizing codes as cosets of linear codes
in order to achieve low encoding and decoding complexity \cite{L,L2}.
However, DSSs appear to have far greater potential and can be exploited to provide self-synchronizing codes with various desired properties.
For our purpose, we would like binary constant-weight self-synchronizing codes of sufficiently large comma-free index with a cartain additional property.
We use DSSs with exactly two sets to obtain desirable codes.

\begin{theorem}\label{DSS}
If there exist a DSS $\{D_0, D_1\}$ of index $\rho$ over ${\textit{\textbf{Z}}}_n$ and
a binary constant-weight code $\mathcal{C}$ of length $n - \vert D_0 \vert - \vert D_1 \vert$ and weight $K$,
then there exists a binary constant-weight self-synchronizing code of length $n$, comma-free index $\rho$, and weight $K+\vert D_1\vert$
with $\vert \mathcal{C} \vert$ codewords.
\end{theorem}
\begin{IEEEproof}
Let $\{D_0, D_1\}$ be a DSS of index $\rho$ over ${\textit{\textbf{Z}}}_n$
and $\mathcal{C}$ a binary constant-weight code of length $n - \vert D_0 \vert - \vert D_1 \vert$ and weight $K$.
For every codeword $\boldsymbol{c} \in \mathcal{C}$,
construct the $n$-dimensional vector $\boldsymbol{d}_{\boldsymbol{c}} = (d_0,\dots,d_{n-1})$,
where $d_i = 0$ for all $i \in D_0$, $d_i = 1$ for all $i \in D_1$, and
the $(n - \vert D_0 \vert - \vert D_1 \vert)$-dimensional vector $(d_i)$, $i \not\in D_0 \cup D_1$, forms $\boldsymbol{c}$.
Each of the resulting $\vert\mathcal{C}\vert$ elements of $\mathbb{F}_2^n$ is of weight $K + \vert D_1 \vert$.
It suffices to show that the set $\mathcal{D} = \{\boldsymbol{d}_{\boldsymbol{c}} \ \vert \ \boldsymbol{c} \in \mathcal{C}\}$
of these vectors forms a code of comma-free index $\rho$.
Take a pair of not necessarily distinct codewords from $\mathcal{D}$ and form a splice $\boldsymbol{s}$ of length $n$
by concatenating the first $s$ bits of one codeword and the last $n-s$ bits of the other for a positive integer $s \leq n-1$.
Take another not necessarily distinct codeword $\boldsymbol{d}$ from $\mathcal{D}$.
Because $\{D_0, D_1\}$ forms a DSS of index $\rho$,
there are at least $\rho$ ordered pairs $(a, b)$ such that $a - b \equiv s \pmod{n}$, where $a$ and $b$ belong to different sets of the DSS.
Thus, there are at least $\rho$ discrepancies between $\boldsymbol{s}$ and $\boldsymbol{d}$ within the coordinates $i \in D_0 \cup D_1$.
The proof is complete.
\end{IEEEproof}

To make the virtue of the above construction clearer,
take the set $\{\{1,2,3,4,5\}, \{0,6,11,16,21\}\}$, which forms a perfect regular DSS of index two over $\textit{\textbf{Z}}_{26}$.
The two sets of cardinality five specify the positions of $0$s and $1$s respectively as synchronization markers
while the remaining $16$ positions are freely available for signal modulation
by a binary constant-weight code of length $16$ such as multipulse PPM with $Q = 16$.
If we use $\{1,2,3,4,5\}$ for $0$s and $\{0,6,11,16,21\}$ for $1$s,
by writing a bit used by the constant-weight code of length $16$ as $*$,
we have 26 bit sequence
\[1000001{*}{*}{*}{*}1{*}{*}{*}{*}1{*}{*}{*}{*}1{*}{*}{*}{*}.\]
Because each nonzero outer difference appears twice in the DSS,
regardless of the content of each $*$,
there are at least two discrepancies among the positions $\{1,2,3,4,5\}\cup\{0,6,11,16,21\}$
between any pair of a valid codeword of the resulting self-synchronizing code and a splice.
To give a smaller example,
we can combine the DSS $\{\{1,2\},\{3,5\}\}$ over ${\textit{\textbf{Z}}}_8$ and a binary constant-weight code of length four such as PPM with $Q = 4$ in the same way.
In this case, the coordinates $0$, $4$, $6$, and $7$ correspond to free bits.
Note that in each of these examples the cardinalities $\vert D_0 \vert$ and $\vert D_1\vert$ are the same.
Hence, we may swap the roles of the sets to obtain a binary constant-weight self-synchronizing code of the same length
and comma-free index guaranteed by Theorem \ref{DSS} which has the same number of codewords.

A self-synchronizing code constructed by the method given in Theorem \ref{DSS} can be synchronized
by only looking at periodic autocorrelations of the partial window specified by the corresponding DSS.
Hence, the synchronization device on the receiver side only needs the corresponding signals as inputs for synchronization.
If we use symbol intervals of PPM, multipulse PPM or expurgated PPM as the binary constant-weight code $\mathcal{C}$
of length $Q = n - \vert D_0 \vert - \vert D_1 \vert$ in Theorem \ref{DSS},
we achieve symbol synchronization that is securely checked for each symbol interval
while allowing for modulation by pulse positions in the time domain through freely available bits.
Packing $f$ symbol intervals into the freely available part provides frame synchronization that is constantly checked for each frame consisting of $f$ symbol intervals.
If we use a DSS over ${\textit{\textbf{Z}}}_n$ for symbol synchronization and then employ another DSS $\{D_0', D_1'\}$ over ${\textit{\textbf{Z}}}_{n'}$
such that $n' - \vert D_0' \vert - \vert D_1' \vert$ is a multiple $f=nf'$ of $n$,
then the scheme provides both symbol and frame synchronization.

In the remainder of this subsection, we explore properties of DSSs
and demonstrate that our synchronization method is significantly more efficient than the known technique
that uses a sequence of symbols as a synchronization marker.

Because a DSS $\{D_0, D_1\}$ over ${\textit{\textbf{Z}}}_n$ leaves $n - \vert D_0 \vert - \vert D_1 \vert$ bits for the modulation layer,
ceteris paribus, it is desirable for $\vert D_0 \vert + \vert D_1 \vert$ to be small.
This parameter is called the \textit{redundancy} of a DSS.
In our context, redundancy is the parameter that denotes the number of time slots we sacrifice
for synchronization per symbol interval or per frame interval.
We use $r(n,\rho)$ to denote the smallest achievable redundancy for given order $n$ and index $\rho$.
A DSS is \textit{optimal} if its redundancy is $r(n,\rho)$.
The following is a special case of the well-known Levenshtein bound:

\begin{theorem}[\cite{L}]\label{lbound}
For any DSS of index $\rho$ over ${\textit{\textbf{Z}}}_n$ with exactly two sets, it holds that
\[r(n,\rho) \geq \sqrt{2\rho(n-1)}\]
with equality if and only if the DSS is perfect and regular.
\end{theorem}

If we are allowed to have a sufficiently strong signal for each pulse compared to noise,
it may be enough to employ a DSS of index one or two.
In this case, the following classical results give optimal DSSs:

\begin{theorem}[\cite{L,C}]\label{rho1}
For any integer $n \geq 2$,
the pair of sets
\begin{align*}D_0 &= \{i\tau_0+1 \ \vert \ 1 \leq i \leq \tau_1\}\ \text{and}\\
D_1 &= \{i \ \vert \ 1 \leq i \leq \tau_0\}
\end{align*}
form an optimal DSS of index one over ${\textit{\textbf{Z}}}_n$, where
\[\tau_0 = \left\lceil\frac{n-1}{2\tau_1}\right\rceil\ \text{and}\ \tau_1 = \left\lceil\sqrt{\frac{n-1}{2}}\right\rceil.\]
\end{theorem}
Note that when the order of the ring ${\textit{\textbf{Z}}}_n$ is of the form $n = 2m^2+1$ for some positive integer $m$,
the redundancy of the optimal DSS given above achieves the Levenshtein bound with equality.
\begin{theorem}[\cite{L}]\label{rho2}
For any integer $n \geq 2$,
the pair of sets
\begin{align*}D_0 &= \{\tau_0+1\}\cup\{n-i\tau_0 \ \vert \ 0 \leq i \leq \tau_1-2\}\ \text{and}\\
D_1 &= \{i \ \vert \ 1 \leq i \leq \tau_0\}
\end{align*}
form an optimal DSS of index two over ${\textit{\textbf{Z}}}_n$, where
\[\tau_0 = \left\lceil\frac{n-1}{\tau_1}\right\rceil\ \text{and}\ \tau_1 = \left\lceil\sqrt{n-1}\right\rceil.\]
\end{theorem}
As in Theorem \ref{rho1}, the redundancies of the optimal DSSs in Theorem \ref{rho2} achieve the lower bound in Theorem \ref{lbound} with equality
when $n = m^2+1$ for some positive integer $m$.

It is notable that,
in terms of the asymptotic notation,\footnote{Here we use the family of Bachmann-Landau notations
defined in standard textbooks in mathematics and computer science such as \cite[Section 9]{ConMath}. 
The Landau symbol $\mathcal{O}(\cdot)$, which is also known as the big-$O$ symbol, is sometimes written
as $O(\cdot)$ with a simple italic $O$ letter in the literature.}
these optimal DSSs only require $\mathcal{O}(n^{\frac{1}{2}})$ bits for synchronization.
For instance, if we use an optimal DSS for symbol synchronization with PPM,
because $n = \mathcal{O}(Q)$, the number of time slots we sacrifice is $\mathcal{O}(Q^{\frac{1}{2}})$ per symbol interval.
Any method that inserts a sequence of valid symbol intervals
must sacrifice $sQ = \Omega(Q)$ time slots for some positive integer $s$,
occupying a significantly larger number of time slots than our method.
It is also worth noting that Theorems \ref{rho1} and \ref{rho2} explicitly give optimal examples for all nontrivial order $n \geq 2$.

If the peak power is severely limited, we may need a DSS of larger index for secure synchronization.
Such DSSs have been studied in various contexts (see, for example, \cite{FV,FL,ZTWY,LF,CLY,D,FMT,CD,T} and references therein).
To study the use of a DSS $\{D_0, D_1\}$ over ${\textit{\textbf{Z}}}_n$ for synchronization in the context of signal modulation by pulse positions,
we use the \textit{redundancy rate} $R = \frac{\vert D_0 \vert + \vert D_1 \vert}{n}$ of the DSS to measure its slot usage. 
For instance, if the redundancy rate is half, synchronization and modulation require the same amount of time resources.
The least useful DSSs for our purpose are those of redundancy rate one because there would be no time slots for modulation.
We aim for the smallest possible redundancy rate for given $n$ and $\rho$.
Note that if we define an analogous parameter for a synchronization method that inserts a sequence of valid codewords for symbol synchronization
by taking the fraction between the numbers of time slots for synchronization and those for modulation per symbol,
such a value can never be less than a half because one ought to insert at least one symbol per symbol interval.
As we will see next, our method can break this fundamental limit even when the channel is assumed to be noisy.

The lower bound on the achievable redundancy given in Theorem \ref{lbound}
suggests that the number of time slots required for synchronization may still be only $\mathcal{O}(n^{\frac{1}{2}})$ for the case when the index is larger than two.
In fact, there are known classes of optimal DSSs with exactly two sets that meet the Levenshtein bound with equality.
In the remainder of this subsection, we list known optimal DSSs that are useful for our purpose
as well as almost optimal DSSs that are equally of interest but are not found in the literature.

Let $p = eg+1$ be an odd prime for some positive integers $e$ and $g$.
The $e$th \textit{cyclotomic classes} in $\mathbb{F}_p$ are defined as
$C_i^e = \{\alpha^{i+te} \ \vert \ 0 \leq t \leq g-1\}$, where $\alpha$ is a primitive element of $\mathbb{F}_p$
and $0 \leq i \leq e-1$.
The \textit{cyclotomic numbers} $(i,j)_e$ of \textit{order} $e$ are defined as $(i,j)_e = \left\vert (C_i^e+1) \cup C_j^e\right\vert$.
Note that the ring ${\textit{\textbf{Z}}}_p$ may be regarded as the finite field $\mathbb{F}_p$ by defining the natural division when $p$ is prime.
We use the following special case of the construction for DSSs given in \cite{MT}:

\begin{theorem}[\cite{MT}]\label{cyclotomic}
Let $p = 2eh+1$ be an odd prime, where $e$ and $h$ are positive integers.
The set $\{C_{0}^{2e}, C_{e}^{2e}\}$ of two cyclotomic classes in $\mathbb{F}_p$ forms
a regular \textup{DSS} over ${\textit{\textbf{Z}}}_p$ and index $\rho$, where
\[\rho = \min\left\{(i, e)_{2e} + (i+e, e)_{2e}\ \middle\vert \ 0 \leq i \leq e-1\right\}.\]
In particular, if
\[(i, e)_{2e} + (i+e, e)_{2e} = \frac{h}{e}\]
for every $i$, then the regular \textup{DSS} is of index $\frac{h}{e}$, perfect, and hence optimal.
\end{theorem}

The following are the two known classes of optimal DSSs with exactly two sets constructed in this manner:
\begin{theorem}[\cite{FV}]\label{half1}
For every $m$ such that $16m^2+1$ is an odd prime,
the set $\{C_0^4, C_2^4\}$ of two cyclotomic classes in $\mathbb{F}_{16m^2+1}$ forms
a perfect regular \textup{DSS} of index $2m^2$ and redundancy rate $\frac{1}{2}-\frac{1}{32m^2+2}$ over ${\textit{\textbf{Z}}}_{16m^2+1}$.
\end{theorem}
\begin{theorem}[\cite{FV}]\label{third}
For every $m$ such that $108m^2+1$ is an odd prime,
the set $\{C_0^6, C_3^6\}$ of two cyclotomic classes in $\mathbb{F}_{108m^2+1}$ forms
a perfect regular \textup{DSS} of index $6m^2$ and redundancy rate $\frac{1}{3}-\frac{1}{324m^2+3}$ over  ${\textit{\textbf{Z}}}_{108m^2+1}$.
\end{theorem}

Because the above optimal DSSs are both perfect and regular at the same time,
their redundancies meet the Levenshtein bound with equality.
For instance, the DSS over ${\textit{\textbf{Z}}}_{16m^2+1}$ in Theorem \ref{half1} uses $8m^2$ time slots for synchronization
and leaves the remaining $8m^2+1$ time slots for modulation.
For comparison to the method that inserts a sequence of symbols,
in the case of symbol synchronization,
this means that the efficiency of the DSSs in Theorem \ref{half1} in terms of slot usage is almost the same as inserting only one symbol per symbol interval.
Any method that inserts a symbol sequence in standard PPM or multipulse PPM
can not assure synchronization at this redundancy rate even if the channel is almost noiseless,
whereas our method tolerates up to $\lfloor\frac{2m^2-1}{2}\rfloor = m^2-1$ bit flips for hard-decision decoding
and an equivalent level of noise for soft-decision decoding.
By the same token, Theorem \ref{third} gives DSSs in which
the number of time slots per symbol interval for synchronization is only about a half of the number of time slots for modulation.
As their redundancy rate $R \approx \frac{1}{3}$ suggests, this level of efficiency in terms of the number of sacrificed time slots is fundamentally unachievable
by any method that inserts valid codewords for synchronization.
Table \ref{numerical} lists the perfect regular DSSs obtained by Theorems \ref{half1} and \ref{third} for $m \leq 10$.
\begin{table}
\renewcommand{\arraystretch}{1.3}
\caption{Perfect regular DSSs from cyclotomic constructions for $m \leq 10$}
\label{numerical}
\centering
\begin{tabular}{cccccc}
\hline\hline
$m$ & $n$ & $\vert D_i \vert$ & $\rho$ & $R = \frac{\vert D_0 \vert + \vert D_1 \vert}{n}$ & \bfseries Reference\\
\hline
$1$ & $17$ & $4$ & $2$ & $\frac{8}{17}$ & Theorem \ref{half1}\\
$4$ & $257$ & $64$ & $32$ & $\frac{128}{257}$ & Theorem \ref{half1}\\
$5$ & $401$ & $100$ & $50$ & $\frac{200}{401}$ & Theorem \ref{half1}\\
$6$ & $577$ & $144$ & $72$ & $\frac{288}{577}$ & Theorem \ref{half1}\\
$9$ & $1297$ & $324$ & $162$ & $\frac{648}{1297}$ & Theorem \ref{half1}\\
$10$ & $1601$ & $400$ & $200$ & $\frac{800}{1601}$ & Theorem \ref{half1}\\
\hline
$1$ & $109$ & $18$ & $6$ & $\frac{36}{109}$ & Theorem \ref{third}\\
$2$ & $433$ & $72$ & $24$ & $\frac{144}{433}$ & Theorem \ref{third}\\
$6$ & $3889$ & $648$ & $216$ & $\frac{1296}{3889}$ & Theorem \ref{third}\\
 \hline
 \hline
\end{tabular}
\end{table}

If we allow the redundancy of a DSS to be slightly above the right-hand side of the lower bound in Theorem \ref{lbound},
we may extend the range of parameters covered by the same construction method while keeping redundancy very low.
This idea was investigated in a more general setting by Mutoh in his unpublished manuscript \cite{Mutoh}.
Here we give two classes of DSSs by simply plugging cyclotomic numbers calculated in \cite{Dickson} into Theorem \ref{cyclotomic}.
For details of the calculations of cyclotomic numbers, we refer the reader to \cite{Dickson,Storer}.
\begin{theorem}\label{4n+1}
Let $n \equiv 1 \pmod{4}$ be a prime and binary quadratic form $n = x^2+4y^2$ with $x \equiv 1 \pmod{4}$ its decomposition.
Then the set $\{C_0^4, C_2^4\}$ of two cyclotomic classes in $\mathbb{F}_n$ forms
a regular \textup{DSS} of index $\rho$ and redundancy rate $\frac{1}{2}-\frac{1}{2n}$ over ${\textit{\textbf{Z}}}_n$, where
\[\rho = 
\begin{cases}
\min\left(\frac{n-3+2x}{8}, \frac{n+1-2x}{8}\right) & \text{if}\ n \equiv 1 \pmod{8},\\
\min\left(\frac{n-3-2x}{8}, \frac{n+1+2x}{8}\right) & \text{otherwise}.
\end{cases}\]
\end{theorem}
\begin{IEEEproof}
Let $n = 4h+1$ be a prime for some integer $h$ and $x^2+4y^2$ with $x \equiv 1 \pmod{4}$ its decomposition.
Take the set $\{C_0^4, C_2^4\}$ of two cyclotomic classes in $\mathbb{F}_n$.
We compute the index $\rho$ of this set as a DSS over ${\textit{\textbf{Z}}}_{n}$.
By Theorem \ref{cyclotomic}, we have $\rho = \min((0,2)_4+(2,2)_4, (1,2)_4+(3,2)_4)$.
If $n \equiv 1 \pmod{8}$, by plugging the actual values of the cyclotomic numbers of order four \cite{Dickson},
we have
\begin{align*}
(0,2)_4+(2,2)_4 &= 2(0,2)_4\\
&= \frac{n-3+2x}{8}
\end{align*}
and
\begin{align*}
(1,2)_4+(3,2)_4 &= 2(1,2)_4\\
&= \frac{n+1-2x}{8}.
\end{align*}
Similarly, if $n \equiv 5 \pmod{8}$, we have
\[(0,2)_4+(2,2)_4 = \frac{n-3-2x}{8}\]
and
\[(1,2)_4+(3,2)_4 = \frac{n+1+2x}{8}.\]
Each cyclotomic class contains $h = \frac{n-1}{4}$ elements of $\mathbb{F}_n$.
Hence, $\{C_0^4, C_2^4\}$ forms a DSS of desired parameters.
\end{IEEEproof}
\begin{theorem}\label{6n+1}
Let $n \equiv 1 \pmod{6}$ be a prime and binary quadratic form $n = x^2+3y^2$ with $x \equiv 1 \pmod{3}$ its decomposition.
Then the set $\{C_0^6, C_3^6\}$ of two cyclotomic classes in $\mathbb{F}_n$
forms a regular \textup{DSS} of index $\rho$ and redundancy rate $\frac{1}{3}-\frac{1}{3n}$ over ${\textit{\textbf{Z}}}_n$, where
\[\rho = 
\begin{cases}
\min\left(\frac{n-5+4x}{18}, \frac{n+1-2x}{18}\right)\\ \quad \text{if}\ 2\ \text{is a cubic residue modulo $n$},\\
\min\left(\frac{n-5+4x+6y}{18}, \frac{n+1-2x-12y}{18}, \frac{n+1-2x+6y}{18}\right)\\ \quad \text{otherwise}.
\end{cases}\]
\end{theorem}
\begin{IEEEproof}
Let $n = 6h+1$ be a prime for some integer $h$ and $x^2+3y^2$ with $x \equiv 1 \pmod{3}$ its decomposition.
Take the set $\{C_0^6, C_3^6\}$ of two cyclotomic classes in $\mathbb{F}_n$
to construct a DSS of index $\rho$ over ${\textit{\textbf{Z}}}_{n}$.
By Theorem \ref{cyclotomic}, we have
\[\rho = \min((0,3)_6+(3,3)_6, (1,3)_6+(4,3)_6, (2,3)_6+(5,3)_6).\]
As in the proof of Theorem \ref{4n+1}, a routine computation shows that
\[
(0,3)_6+(3,3)_6 =
\begin{cases}
\frac{n-5+4x}{18} & \text{if}\ 2\ \text{is a cubic residue},\\
\frac{n-5+4x+6y}{18} & \text{otherwise},
\end{cases}
\]
that
\[
(1,3)_6+(4,3)_6 =
\begin{cases}
\frac{n+1-2x}{18} & \text{if}\ 2\ \text{is a cubic residue},\\
\frac{n+1-2x-12y}{18} & \text{otherwise},
\end{cases}
\]
and that
\[
(2,3)_6+(5,3) =
\begin{cases}
\frac{n+1-2x}{18} & \text{if}\ 2\ \text{is a cubic residue},\\
\frac{n+1-2x+6y}{18} & \text{otherwise}.
\end{cases}
\]
Each cyclotomic class contains $h = \frac{n-1}{6}$ elements of $\mathbb{F}_n$.
Hence, we obtain a DSS of desired parameters.
\end{IEEEproof}

Note that Theorems \ref{half1} and \ref{third} are special cases of these two classes
in which the resulting DSSs are simultaneously perfect and regular.
In general, Theorem \ref{cyclotomic} produces an optimal DSS or one close to optimal
when the cyclotomic classes in $\mathbb{F}_p$ can be taken such that
$(i, e)_{2e} + (i+e, e)_{2e}$ are uniform or almost uniform across $i$.
For instance, take $n = 37 = 1^2+4\cdot3^2$. Then by Theorem \ref{4n+1},
we obtain a DSS of index $4$ and redundancy 18 over ${\textit{\textbf{Z}}}_{37}$.
By the Levenshtein bound, the redundancy of any DSS of the same index with exactly two sets over ${\textit{\textbf{Z}}}_{37}$
must be at least
\[\left\lceil \sqrt{2\cdot4\cdot(37-1)} \right\rceil = 17,\]
which is very close to $18$.

If lower time slot usage is desirable,
in principle, regular DSSs of better redundancy rate can be obtained in the same way in exchange for poorer indices $\rho$
by applying cyclotomic numbers of higher orders.
If higher indices are required to tolerate a higher noise level,
such perfect DSSs can be constructed, albeit with more complicated computation,
by taking unions of cyclotomic classes and increasing time slot usage accordingly (see \cite{FMY}).

DSSs of index larger than two can be obtained by known recursive constructions as well.
The following is a relevant special case of the recursive constructions given in \cite{FL}.
\begin{theorem}[\cite{FL}]\label{recursive}
Let $n = \frac{q^{t+1}-1}{q-1}$ and $n' = \frac{q^{2t+2}-1}{q-1}$,
where $q$ is a prime power and $t$ a positive integer.
If there exists a \textup{DSS} $\{D_0, D_1\}$ of index $\rho$ over ${\textit{\textbf{Z}}}_n$,
then there exist a \textup{DSS} $\{D'_0, D'_1\}$ of index $\rho'$ over ${\textit{\textbf{Z}}}_{n'}$, where
\[\vert D'_i \vert = q^{t+1}\vert D_i\vert \ \text{for}\ i = 0, 1\]
and
\[\rho' = \min\left(\rho q^{t+1}, 2(q-1)\vert D_0\vert\vert D_1\vert\right),\]
and a \textup{DSS} $\{D''_0, D''_1\}$ of index $\rho''$ over ${\textit{\textbf{Z}}}_{n'}$, where
\[\vert D''_i \vert = q^{t+1}\vert D_i\vert + in \ \text{for}\ i = 0, 1\]
and
\[\rho'' = \min\left(\rho q^{t+1}, 2(q-1)\vert D_0\vert\vert D_1\vert + 2\vert D_0 \vert\right).\]
\end{theorem}
This recursive construction gives DSSs of improved index by increasing the time slot usage for synchronization.
For instance, if we apply the first half of Theorem \ref{recursive} to the optimal DSS of index one over ${\textit{\textbf{Z}}}_n$
with $n = \frac{q^{t+1}-1}{q-1}$ obtained by Theorem \ref{rho1},
the index of the resulting DSS is at least $q^{t+1}-q$ because we have
\begin{align*}
2(q-1)\tau_0\tau_1 &\geq 2(q-1)\tau_1^2\\
&\geq q^{t+1}-q.
\end{align*}

While we have focused on theoretical aspects of DSSs and systematic constructions,
one may also look for optimal DSSs of specific parameters through computer searches.
An algorithm for finding optimal DSSs is proposed in \cite{TW}.
An explicit example of optimal DSSs of index $\rho$ over ${\textit{\textbf{Z}}}_n$ for each $\rho \leq 5$ and $n \leq 30$
can be found in \cite{web}.
The computer search results and the existence of systematic constructions for particular parameters seem to suggest that
while it is quite difficult to give explicit constructions,
the redundancies of optimal DSSs of index $\rho$ over ${\textit{\textbf{Z}}}_n$ are generally very close or equal to $\sqrt{2\rho(n-1)}$.

\subsection{Modulation layer}\label{mod}
We now turn our attention to the modulation layer of our scheme.
As we have seen in the previous subsection,
we can employ standard PPM and its variations by exploiting the freely available bits given by a DSS.
One major benefit of using a DSS is that it allows for error tolerant synchronization
while still keeping the number of time slots for synchronization per symbol very low.
One might then wish error correction for the modulation layer at the modulation stage
to eliminate the need of or reduce the burden on the shoulders of error correction at a higher level.

Expurgated PPM is the error-correcting variant of PPM which can be understood as a binary constant-weight error-correcting code.
To take advantage of our coding theoretic framework, we first briefly review this modulation technique and describe it
in the language of constant-weight codes.
A more general error-correcting variant of expurgated PPM will then be developed to accommodate a larger number of symbols.

Expurgated PPM employs special combinatorial designs with cyclic automorphisms.
A \textit{simple} $2$-\textit{design} of \textit{order} $v$, \textit{block size} $k$, and \textit{index} $\mu$
is an ordered pair $(V, \mathcal{B})$, where $V$ is a finite set of cardinality $v$
and $\mathcal{B}$ is a set of $k$-subsets of $V$ such that each pair of elements of $V$ is included in exactly $\mu$ elements of $\mathcal{B}$.
Elements of $V$ are called \textit{points} while those of $\mathcal{B}$ are \textit{blocks}.
A simple $2$-design $(V, \mathcal{B})$ of order $v$ is said to be \textit{symmetric} if $\vert \mathcal{B} \vert = \vert V \vert = v$.
It is \textit{cyclic} if the cyclic group of order $v$ acts regularly on the points.

A \textit{difference set} of order $v$ and index $\mu$ is a set $B$ of non-negative integers less than $v$
such that every element of $\textit{\textbf{Z}}_v \setminus \{0\}$ appears exactly $\mu$ times each
as the difference $a - b \pmod{v}$ between two distinct elements $a, b \in B, a \not= b$.
To avoid the trivial case, we assume that $k > \mu$.
Let $\pi$ be the map $B \mapsto B+1 = \{b +1 \pmod{v} \ \vert \ b \in B\}$.
It is straightforward to see that if the orbit $\text{\textit{Orb}}_{\textit{\textbf{Z}}_v}(B) = \bigcup_{i \in \textit{\textbf{Z}}_v}\left\{\pi^i(B)\right\}$ is of length $v$,
then the $v$ subsets of $\textit{\textbf{Z}}_v$ form a cyclic simple $2$-design of order $v$ and index $\mu$ that is symmetric.
Its block size is $\vert B \vert = \frac{1+\sqrt{4(v-1)\mu+1}}{2}$.

Expurgated PPM employs the $v$ blocks of a simple $2$-design constructed from a difference set $B$ of order $v$.
Trivially, the block set $\mathcal{B} = \text{\textit{Orb}}_{\textit{\textbf{Z}}_v}(B)$ of this symmetric design forms the set of supports of the codewords
of a binary constant-weight code of length $v$ and weight $k = \frac{1+\sqrt{4(v-1)\mu+1}}{2}$ with $v$ codewords.
Because every nonzero difference appears exactly $\mu$ times in $B$, the minimum distance is $2(k-\mu)$.
Conversely, it is straightforward to see that a binary constant-weight code of these parameters in which
every cyclic shift of a codeword is also a codeword forms a difference set.
By employing this binary constant-weight code for modulation based on pulse positions as in PPM,
we have $Q = v$ time slots for each symbol interval in which
$K = k = \frac{1+\sqrt{4(v-1)\mu+1}}{2}$ pulses are transmitted to represent $M = v = Q$ symbols.
Expurgated PPM thus supports the same number of symbols as standard PPM and has an increased minimum distance.
The code represented by $\mathcal{B}$ has the property that every cyclic shift of a codeword is also a codeword.
Hence, implementation on the receiver side only requires a simple correlation receiver \cite{NB,NB2}.
The above coding theoretic interpretation may be summarized by the following proposition:
\begin{proposition}\label{prop}
The symbols of expurgated PPM that transmits $K$ pulses in each $Q$ slot interval
are equivalent to a binary constant-weight code of length $Q$, weight $K$, and minimum distance $\frac{2K(Q-K)}{Q-1}$ with $Q$ codewords
in which every cyclic shift of a codeword is also a codeword.
\end{proposition}
\begin{IEEEproof}
The set of the $Q$ symbols of expurgated PPM is a subset $\mathcal{P}'_K$ of
the set $\mathcal{P}_K = \left\{\boldsymbol{v}_i \in \mathbb{F}_2^Q \ \middle\vert\ \operatorname{wt}(\boldsymbol{v}_i) = K\right\}$
of all $Q$-dimensional binary vectors $\boldsymbol{v}_i$ of weight $K$ such that
the set $\mathcal{B} = \{\text{supp}(\boldsymbol{v}_i)\ \vert \ \boldsymbol{v}_i \in \mathcal{P}'_K\}$ of supports of the $Q$-dimensional vectors in $\mathcal{P}'_K$
forms the block set of a cyclic simple $2$-design of order $Q$ and block size $K$ that is symmetric.
Let $\mu$ be the index of this corresponding symmetric $2$-design.
It suffices to show that the minimum distance $2(K-\mu)$ of the corresponding constant-weight code is equal to $\frac{2K(Q-K)}{Q-1}$.
Because every pair of points is included in exactly $\mu$ blocks while each of the $Q$ blocks contains ${{K}\choose{2}}$ pairs,
we have
\[\mu{{Q}\choose{2}} = Q{{K}\choose{2}},\]
which implies that
\[\mu = \frac{K(K-1)}{Q-1}.\]
Hence, we have
\begin{align*}
2(K-\mu) &= 2\left(K-\frac{K(K-1)}{Q-1}\right)\\
&= \frac{2K(Q-K)}{Q-1}
\end{align*}
as desired.
\end{IEEEproof}

We generalize expurgated PPM by taking advantage of the above interpretation.
Our approach is to use more general constant-weight codes to realize a variety of parameters of PPM schemes.
For instance, we may increase the number of orbits over $\textit{\textbf{Z}}_v$ while keeping the minimum distance large,
so that the number of codewords becomes a multiple of $v$ rather than exactly $v$.
As we will see in this section, this idea can be formalized through coding theory.

A $(v,k,\lambda)$ \textit{optical orthogonal code}
$\mathcal{C} \subseteq \mathbb{F}_2^v$ of \textit{length} $v$, \textit{weight} $k$, and \textit{index} $\lambda$ is
a set of $v$-dimensional binary vectors of weight $k$ such that
for any $\boldsymbol{c} \in \mathcal{C}$ its off-peak periodic autocorrelations are at most $\lambda$
and for any pair of distinct codewords $\boldsymbol{c}, \boldsymbol{c}' \in \mathcal{C}$ their periodic cross-correlations are at most $\lambda$.
In other words, it is a set of $v$-dimensional vectors with $k$ $1$s and $v-k$ $0$s whose coordinates are indexed by $\textit{\textbf{Z}}_v$ such that
\[\sum_{0 \leq t \leq v-1}c_t c_{t+i} \leq \lambda\]
for any $\boldsymbol{c} = (c_0, c_1, \dots, c_{v-1}) \in \mathcal{C}$ and any nonzero element $i \in \textit{\textbf{Z}}_v$ and such that
\[\sum_{0 \leq t \leq v-1}c_t c'_{t+i} \leq \lambda\]
for any pair of distinct vectors $\boldsymbol{c} = (c_0, c_1, \dots, c_{v-1}),  \boldsymbol{c}' = (c'_0, c'_1, \dots, c'_{v-1})\in \mathcal{C}$
and any $i \in \textit{\textbf{Z}}_v$.
We do not consider the trivial case $k = \lambda$ and always assume that $k > \lambda$.
We allow the special case $\vert \mathcal{C} \vert = 1$ as long as the autocorrelation property holds for the unique codeword.

Optical orthogonal codes have been extensively investigated from various viewpoints including
the initial motivation in the context of code-division multiple-access fiber optical communications \cite{CSW}.
A useful observation for our goal is that an optical orthogonal code is equivalent to
a binary constant-weight code in which every cyclic shift of a codeword is also a distinct codeword,
which is the property we would like for our signal modulation purpose.
To see the equivalence, take the union $\mathcal{D}$
of a $(v,k,\lambda)$ optical orthogonal code $\mathcal{C}$ and the set of all $v-1$ distinct cyclic shifts of each codeword.
Then $\mathcal{D}$ forms a binary constant-weight code of length $v$, weight $k$, and minimum distance $2(k-\lambda)$.
Trivially, the converse also holds.
Because we included all cyclic shifts of $\boldsymbol{c} \in \mathcal{C}$ in $\mathcal{D}$, 
every cyclic shift of $\boldsymbol{d} \in \mathcal{D}$ is naturally in $\mathcal{D}$ again.
Because the condition that $k > \lambda$ is assumed,
the number of codewords of the corresponding binary constant-weight code is $v\vert \mathcal{C} \vert$.

Now by Proposition \ref{prop},
if we look at expurgated PPM with $v$ symbols that uses $k$ pulses per symbol in our coding theoretic framework,
it is a binary constant-weight code of length $v$, weight $k$, and minimum distance $2(k-\lambda)$
with exactly $v$ codewords in which every cyclic shift of a codeword is also a codeword.
In other words, it is simply the union of a special $(v,k,\lambda)$ optical orthogonal code with only one codeword and its $v-1$ cyclic shifts.
Because it is also a symmetric design, we have $\lambda = \frac{k(k-1)}{v-1}$.
Thus, we obtain the following proposition:
\begin{proposition}
The set of symbols of expurgated PPM that transmits $K$ pulses in each $Q$-slot interval is equivalent to
a $(Q, K, \frac{K(K-1)}{Q-1})$ optical orthogonal code with exactly one codeword.
\end{proposition}

Since an optical orthogonal code $\mathcal{C}$ of length $v$ gives rise to a binary constant-weight code with $v\vert\mathcal{C}\vert$ codewords
by joining the $v-1$ distinct cyclic shifts of all $\boldsymbol{c} \in \mathcal{C}$,
it is natural to generalize the PPM method such that modulation exploits an optical orthogonal code with more than one codeword.
With this generalization, a larger number of symbols can be supported compared to expurgated PPM
while still using the same decoder for each orbit and maintaining the error correction mechanism at the modulation stage.
Summarizing the discussion given above in this subsection, we have the following theorem:
\begin{theorem}\label{generalPPM}
The set of the codewords of a $(v,k,\lambda)$ optical orthogonal code $\mathcal{C}$ and all cyclic shifts of each codeword
defines a $(v\vert\mathcal{C}\vert)$-ary signal modulation technique
in which each of the $v\vert\mathcal{C}\vert$ symbols with mutual Hamming distance at least $2(k-\lambda)$
is represented by $k$ single pulses transmitted at $k$ out of $v$ time slots.
\end{theorem}
It is notable that this generalization of expurgated PPM also allows for the case when the code is of size $\vert \mathcal{C} \vert = 1$
but does not form a difference set. An example of this type of optical orthogonal code is the binary vector representation of
an \textit{almost difference set} given in \cite{DHL}.
Because difference sets are known to be difficult to construct,
allowing optical orthogonal codes that are not difference sets
substantially extends the range of possible parameters of modulation schemes of PPM type.

We would like optical orthogonal codes with the largest possible number of codewords for given $v$, $k$, and $\lambda$.
Because an optical orthogonal code $\mathcal{C}$ is already a constant-weight code before joining the cyclic shifts of each codeword,
the Johnson bound gives the upper bound on the number of codewords of $\mathcal{C}$:
\begin{theorem}[\cite{Johnson}]\label{jb}
Let $\mathcal{C}$ be a $(v,k,\lambda)$ optical orthogonal code.
Then it holds that
\[
\vert \mathcal{C} \vert \leq
\left\lfloor\frac{1}{k}\left\lfloor\frac{v-1}{k-1}\left\lfloor\frac{v-2}{k-2}\left\lfloor\cdots\left\lfloor\frac{v-\lambda}{k-\lambda}
\right\rfloor\cdots\right\rfloor\right\rfloor\right\rfloor\right\rfloor.
\]
\end{theorem}

A $(v,k,\lambda)$ optical orthogonal code is \textit{optimal} if the number of codewords attains this upper bound.
Table \ref{ooctable} lists well-known classes of optimal optical orthogonal codes of index one with more than one codeword that span a variety of lengths and weights.
\begin{table*}
\renewcommand{\arraystretch}{1.6}
\caption{Some classes of optimal optical orthogonal codes of index one with more than one codeword}
\label{ooctable}
\centering
\begin{tabular}{cccccc}
\hline\hline
\bfseries Length $v$ & \bfseries Weight $k$ & \bfseries Index $\lambda$ & \bfseries Number $\vert \mathcal{C} \vert$ of codewords & \bfseries Constraint & \bfseries Reference\\
\hline
$n$ & $3$ & $1$  & $\left\lfloor\frac{n-1}{6}\right\rfloor$ & $n\not\equiv 14, 20 \pmod{24}$\rlap{\textsuperscript{a}} & \cite{Peltesohn,CSW} \\
$n$ & $4$ & $1$  & $\left\lfloor\frac{n-1}{12}\right\rfloor$ & $n \equiv 0, 6, 18 \pmod{24}$ & \cite{GY,CFM,CM} \\
\multirow{3}{*}{$p$} & \multirow{3}{*}{any integer $k$} & \multirow{3}{*}{$1$}  & \multirow{3}{*}{$\frac{p-1}{k(k-1)}$} & $p \equiv 1 \pmod{k(k-1)}$ is prime, & \multirow{3}{*}{\cite{WilsonC}} \\
 & &  & & $p > c_k$, &\\
 & &  & & $c_k$ is a constant dependent on $k$\rlap{\textsuperscript{b}} &\\
$q^t-1$ & $q$ & $1$  & $\frac{q^{t-1}-1}{q-1}$ & $q$ is a prime power& Affine geometry with origine deleted\rlap{\textsuperscript{c}}\\
$\frac{q^{t+1}-1}{q-1}$ & $q+1$ & $1$ & $\begin{cases}\frac{q^t-1}{q^2-1}, \ t\ \text{even}\\ \frac{q^t-1}{q^2-1}, \ t\ \text{odd}\end{cases}$ & $q$ is a prime power& Projective geometry\\
 \hline
 \hline
\multicolumn{6}{l}{\scriptsize\textsuperscript{a}
This is a necessary and sufficient condition for the existence (see \cite{AB} for a short proof).
The constrains for the other classes are sufficient conditions.}\vspace{-1.1mm}\\
\multicolumn{6}{l}{\scriptsize\textsuperscript{b} $c_4 = c_5 = 0$ \cite{CZ}.
$c_6 = 61$ \cite{CZ2}. For $k \geq 7$ in general, the best known value is $c_k = {{k}\choose{2}}^{k(k-1)}$ \cite{WilsonC}.}\vspace{-1.1mm}\\
\multicolumn{6}{l}{\scriptsize\textsuperscript{c} The same parameters may be realized as a generalized Bose-Chowla family \cite{MOKL}.}\vspace{2.2mm}
\end{tabular}
\end{table*}
There are numerous other constructions and existence results.
For the case when the index is one, all other known results can be found in \cite{HandbookCD,WC,RR,BT,Momi,YYL} and references therein.
For the latest results on optical orthogonal codes of higher index, we refer the reader to \cite{HandbookCD,FCJ,FCJ2,FM2} and references therein.

In the remainder of this section, we give some examples of our generalized expurgated PPM
to show how our modulation method enriches the family of PPM schemes.
Because typical signal modulation is $2^m$-ary for some positive integer $m$,
we focus on the case when the number $M$ of symbols is a power of $2$.

To present binary constant-weight codes for modulation in a compact way, codewords are given in terms of their supports.
Take a constant-weight code $\mathcal{D} \subseteq \mathbb{F}_2^Q$ of length $Q$, constant weight $K$,
and minimum distance $2(K-\lambda)$ in which every cyclic shift of every codeword is also a codeword.
Let $\mathcal{B} = \{\operatorname{supp}(\boldsymbol{d}) \ \vert \ \boldsymbol{d} \in \mathcal{D}\}$ be the set of supports of all vectors in $\mathcal{D}$.
Then for any $B \in \mathcal{B}$, we have $B + 1 \in \mathcal{B}$.
In other words, $\mathcal{B}$ is a set of subsets of $V = \{0,1,\dots,Q-1\}$ in which the cyclic group of order $Q$ acts regularly on $V$.
Hence, the elements of $\mathcal{B}$ can be partitioned into orbits
$\text{\textit{Orb}}_{\textit{\textbf{Z}}_Q}(B) = \bigcup_{i \in \textit{\textbf{Z}}_Q}\left\{\pi^i(B)\right\}$, $B \in \mathcal{B}$,
where $\pi(B) = B+1$.
As we have seen in this section, if each orbit is of the same size $Q$, a system of representatives of these orbits forms
the set of supports of all codewords of a $(Q,K,\lambda)$ optical orthogonal code.
Trivially, if $K > \lambda$, the converse also holds.

With this relation, optical orthogonal codes can always be written by finite sets.
For instance, an optimal $(8,3,1)$ optical orthogonal code with a single codeword $(1,1,0,1,0,0,0,0)$ can be understood as a single set $\{0,1,3\}$
because there is $1$ at coordinate $i \in \{0,1,3\}$ and $0$ otherwise.
The binary constant-weight code $\mathcal{D}$ for modulation
in which the positions of $1$s of a codeword represent the positions of pulses in the corresponding symbol
is exactly the set
\begin{align*}
\mathcal{D} = \{&(1,1,0,1,0,0,0,0), (0,1,1,0,1,0,0,0),\\ &(0,0,1,1,0,1,0,0), (0,0,0,1,1,0,1,0),\\ &(0,0,0,0,1,1,0,1), (1,0,0,0,0,1,1,0),\\ &(0,1,0,0,0,0,1,1), (1,0,1,0,0,0,0,1)\}
\end{align*}
obtained by joining the cyclic shifts of $(1,1,0,1,0,0,0,0)$.
If there are two or more codewords in an optical orthogonal code,
then each of the corresponding sets obtained by taking the supports forms a distinct orbit in $\mathcal{D}$.
For more mathematical details of the set representation of an optical orthogonal code, we refer the reader to \cite{FM}.

Table \ref{tablePPM} lists some examples of $M$-ary PPM, MPPM, expurgated PPM, and generalized expurgated PPM for $M = 8$, $16$, and $32$.
\begin{table*}
\renewcommand{\arraystretch}{1.6}
\caption{Small $M$-ary PPM for $M = 2^m$}
\label{tablePPM}
\centering
\begin{tabular}{cccccc}
\hline\hline
\bfseries Type\rlap{\textsuperscript{a}} & \bfseries Number $M$ of Symbols & \bfseries Interval Size $Q$ & \bfseries Number $K$ of Pulses & \bfseries Minimum Distance $d$ & \bfseries Optical Orthogonal Code\rlap{\textsuperscript{b}}\\
\hline
PPM & $8$ & $8$  & $1$ & $2$ & N/A\\
GEPPM & $8$ & $8$  & $3$ & $4$ & $\{0,1,3\}$\\
EPPM & $8$ & $11$  & $5$ & $6$ & $\{0,2,3,4,8\}$\\
\hline
PPM & $16$ & $16$  & $1$ & $2$ & N/A\\
AEPPM & $16$ & $11$  & $5$ & $5$ & $\{0,2,3,4,8\}, \{1,5,6,7,9,10\}$\\
GEPPM & $16$ & $16$  & $4$ & $6$ & $\{0,1,3,7\}$\\
GEPPM & $16$ & $16$  & $8$ & $8$ & Almost difference set \cite[Theorem 4]{ADHKM}\rlap{\textsuperscript{c}}\\
EPPM & $16$ & $19$  & $9$ & $10$ & Paley-type difference set \cite{DSref}\rlap{\textsuperscript{d}}\\
\hline
PPM & $32$ & $32$  & $1$ & $2$ & N/A\\
MPPM & $32$ & $7$  & $3$ & $2$ & N/A\\
GEPPM & $32$ & $16$  & $3$ & $4$ & $\{0,1,3\}, \{0,4,9\}$\\
GEPPM & $32$ & $37$  & $10$ & $14$ & Almost difference set \cite[Theorem 3]{DHL}\rlap{\textsuperscript{e}}\\
EPPM & $32$ & $35$  & $17$ & $18$ & Difference set \cite{DSref}\rlap{\textsuperscript{f}}\\
 \hline
 \hline
\multicolumn{6}{l}{\scriptsize\textsuperscript{a}
This column indicates the type of modulation.
GEPPM stands for our proposed PPM that generalizes expurgated PPM. AEPPM is a variation of EPPM given in \cite{NB}.}\vspace{-1.1mm}\\
\multicolumn{6}{l}{\scriptsize\textsuperscript{b}
Explicit examples are given for small optical orthogonal codes of which the origins are unknown.}\vspace{-1.1mm}\\
\multicolumn{6}{l}{\scriptsize\textsuperscript{c}
The $(n,k,\lambda,t)$ almost difference sets given in \cite[Theorem 4]{ADHKM} are
$(n,k,\lambda+1)$ optical orthogonal codes of size $\vert \mathcal{C}\vert = 1$.}\vspace{-1.1mm}\\
\multicolumn{6}{l}{\scriptsize\textsuperscript{d}
An example of its set representation is $\{0,3,4,5,6,8,10,15,16\}$.}\vspace{-1.1mm}\\
\multicolumn{6}{l}{\scriptsize\textsuperscript{e}
The $(n,k,\lambda)$ almost difference sets given in \cite[Theorem 3]{DHL} are
$(n,k,\lambda+1)$ optical orthogonal codes of size $\vert \mathcal{C}\vert = 1$.}\vspace{-1.1mm}\\
\multicolumn{6}{l}{\scriptsize\textsuperscript{f} An example of its set representation is $\{0,1,3,4,7,9,11,12,13,14,16,17,21,27,28,29,33\}$.}\vspace{5.5mm}
\end{tabular}
\end{table*}
To keep the table concise, the set representations of example optimal optical orthogonal codes
for expurgated PPM and its generalized versions are given only when the origins seem unknown.
For other cases, references are given instead.

As is shown in the table, our modulation scheme greatly widens the range of available parameters of modulation techniques of PPM type.
For instance, when compared to standard $8$-ary PPM,
$8$-ary EPPM achieves large minimum distance $6$ by increasing the interval size from $8$ to $11$ and the number of pulses from $1$ to $5$.
In other words, $8$-ary EPPM increases the error tolerance capability
by a large extent in exchange for increased energy per symbol and a poorer information rate.
Our $8$-ary modulation based on an $(8,3,1)$ optimal optical orthogonal code is a middle ground approach
in that it does not sacrifice the information rate and only slightly increases the required energy per symbol
in order to achieve good minimum distance.
As is also illustrated by the examples for the $16$-ary and $32$-ary cases in the table,
our generalized scheme typically complements PPM, MPPM, and EPPM by offering a solution that falls between the three extreme approaches.

\section{Concluding remarks}
We introduced a coding theoretic framework to the study of signal modulation based on pulse positions in the time domain.
With the further help of combinatorial design theory,
this approach allowed us to develop a self-synchronizing scheme with significantly improved efficiency
while maintaining compatibility to existence modulation techniques such as standard pulse position modulation.
In fact, the number of time slots required per symbol for synchronization is now improved from $\Omega(Q)$ to $\mathcal{O}(Q^{\frac{1}{2}})$ in the asymptotic sense,
breaking the fundamental limit of the previously known synchronization method.

We were also able to generalize the recently introduced error-correcting pulse position modulation technique to realize a larger number of symbols.
This generalization is particularly appealing as the modulation layer for our synchronization method
because this way error correction can be fully supported at the modulation stage.

In the previous section on synchronization and error correction, we placed particular emphasis on generality so as not to unnecessarily spoil the potential.
For this reason, our focus has been on the minimum distance.
However, it would also be of importance to investigate various other aspects of our scheme by assuming a particular context.
In fact, PPM and its variants have extensively been studied with various applications in mind
(see, for example, \cite{CS,ZC} for use in ultra-wideband communications,
\cite{WBCB} for free-space optics scenarios, and \cite{Ohtsuki} for the purpose of atmospheric optical code-division multiple-access).
Among many possible directions of more focused research, it would be of particular interest
to analyze the efficiency in detail and more accurately estimate the synchronization error rate and bit error rate over a reasonably realistic channel.

If one wishes to extract finer structural information than minimum distance for a detailed analysis in a specific context,
the algebraic properties of DSSs and optical orthogonal codes may be effectively exploited.
For instance, it is straightforward to see that if we employ an optical orthogonal code $\mathcal{C}$ of length $v$, block size $k$, and index $1$
to form the binary constant-weight code $\mathcal{D}$ for modulation by joining cyclic shifts,
for any codeword $\boldsymbol{d} \in \mathcal{D}$ the number $n_{\boldsymbol{d}}$
of codewords of Hamming distance $2(k-1)$ from $\boldsymbol{d}$, which are the nearest,
and the number $f_{\boldsymbol{d}}$ of other codewords except $\boldsymbol{d}$, which are all of Hamming distance $2k$ from $\boldsymbol{d}$, are
\[n_{\boldsymbol{d}} = k^2\vert\mathcal{C}\vert - k\]
and
\begin{align*}
f_{\boldsymbol{d}} &= v\vert\mathcal{C}\vert - 1 - n_{\boldsymbol{d}}\\
&= (v-k^2)\vert\mathcal{C}\vert+k-1
\end{align*}
respectively.

\begin{figure}
\centering
\includegraphics[width=3.3in]{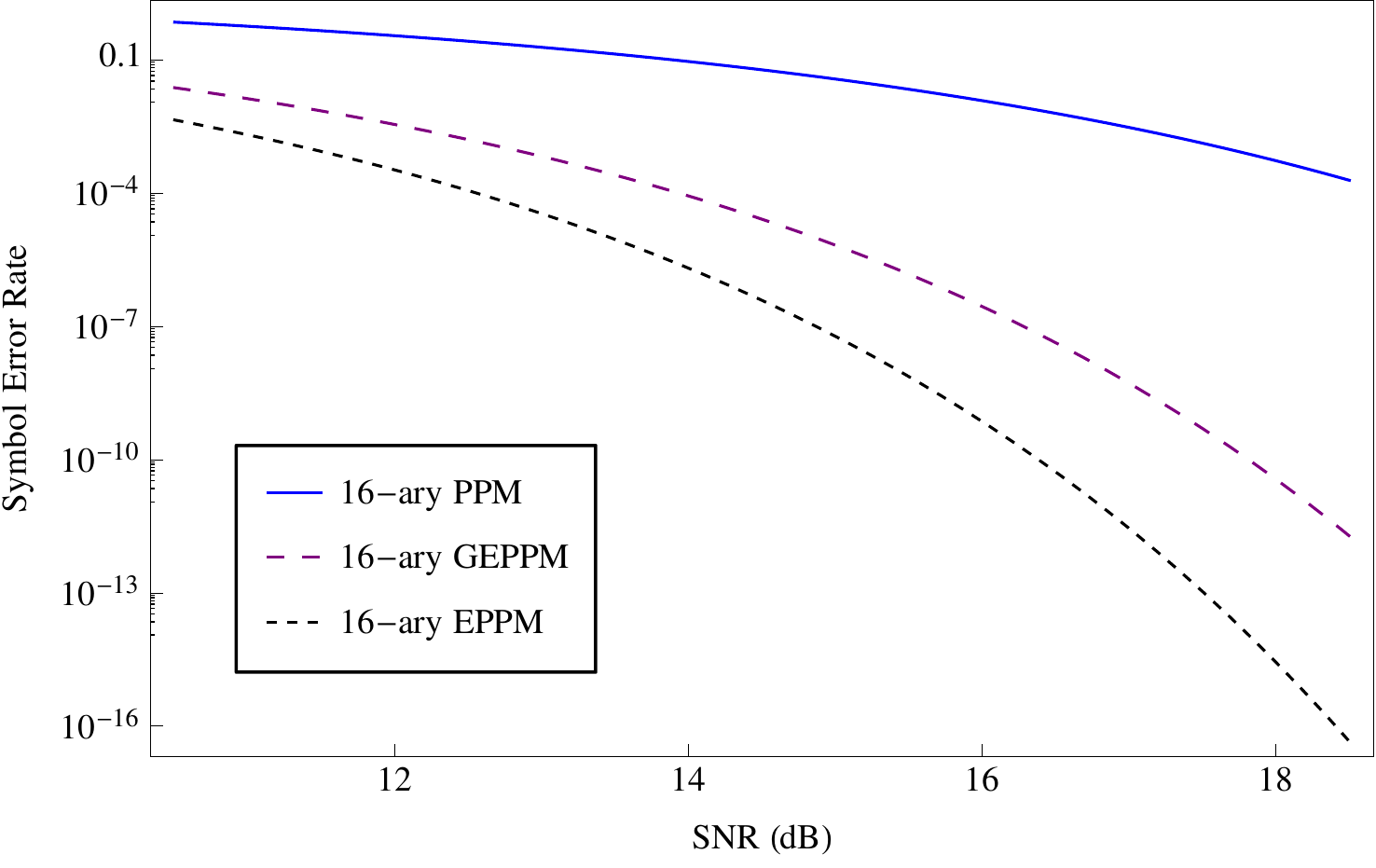}
\caption{Estimated symbol error rates of $16$-ary modulation. $\vert$
PPM, generalized EPPM based on an $(8,4,1)$ optical orthogonal code, and EPPM based on a Paley-type difference set are compared
by the union bound at high SNR.}
\label{figure}
\end{figure}

As an example use of structural information for obtaining a context-specific performance estimation,
assume that pulses are transmitted through a typical free-space optical link
approximated by the AWGN channel with power spectral density, say, $\frac{N_0}{2}$.
Supposing that the decoder is optimal
and that all codewords of a $(v,k,\lambda)$ optical orthogonal code $\vert \mathcal{C} \vert$ and their cyclic shifts are used for modulation,
the symbol error probability $P_s$ of our generalized PPM scheme $\mathcal{D}$ can be estimated by the union bound
\begin{align*}
P_s &\leq \frac{1}{2M}\sum_{\substack{\boldsymbol{d}, \boldsymbol{d}'\in\mathcal{D}\\ \boldsymbol{d}\not=\boldsymbol{d}'}}
\operatorname{erfc}\left(\sqrt{\frac{\gamma\operatorname{wt}(\boldsymbol{d}\oplus\boldsymbol{d}')\log{M}}{2Q}}\right)\\
&=\frac{k^2\vert \mathcal{C} \vert-k}{2}\operatorname{erfc}\left(\sqrt{\frac{\gamma(k-1)\log{(v\vert\mathcal{C}\vert)}}{v}}\right)\\
&\quad + \frac{(v-k^2)\vert\mathcal{C}\vert+k-1}{2}\operatorname{erfc}\left(\sqrt{\frac{\gamma k\log{(v\vert\mathcal{C}\vert)}}{v}}\right),
\end{align*}
where $\operatorname{erfc}$ is the complementary error function, $\oplus$ is the bitwise sum modulo $2$, and
$\gamma = \frac{\rho^2P_0^2}{N_0R_b}$ is the signal-to-noise ratio with $\rho$, $P_0$, and $R_b$ being
the photodetector responsivity, peak optical power, and bit-rate respectively (see \cite{Guimaraes}).
Fig.\ \ref{figure} compares the performance of our $16$-ary error-tolerant PPM based on an $(8,4,1)$ optimal optical orthogonal code with PPM and EPPM.
As expected from their minimum distances, our scheme offers good error tolerance while requiring only a modest amount of energy per symbol.
One may analyze the properties and performance in more detail for a very specific channel in a similar manner as well.

Another aspect we did not address is the relation of our version of expurgated PPM to error-correcting codes at a higher level.
If we see PPM as a binary constant-weight code, the idea of error correction at a higher level is a code concatenation in one sense.
There have been proposed various types of code and their uses for such concatenations
(see, for instance, \cite{McEliece,PB,DVN,TGALF,NL}).
It would be of interest and importance to understand how best to exploit our scheme
along with an error-correcting code at a higher level in a particular communications system.

\section*{Acknowledgment}
Y.F. thanks the three anonymous reviewers
and Associate Editor Robert Fischer for careful reading of the manuscript and constructive suggestions.


\begin{IEEEbiographynophoto}{Yuichiro Fujiwara}
(M'10) received the B.S. and M.S. degrees in mathematics from Keio University, Japan,
and the Ph.D. degree in information science from Nagoya University, Japan.

He was a JSPS postdoctoral research fellow with the Graduate School of
System and Information Engineering, Tsukuba University, Japan, and a visiting
scholar with the Department of Mathematical Sciences, Michigan Technological University.
He is currently with the Division of Physics, Mathematics and Astronomy,
California Institute of Technology, Pasadena, where he works as a JSPS postdoctoral research fellow.

Dr.\ Fujiwara's research interests include combinatorics and its interaction with computer science and quantum information science,
with particular emphasis on combinatorial design theory, algebraic coding theory, and quantum information theory.
\end{IEEEbiographynophoto}

\end{document}